\documentclass[12pt]{iopart}

\usepackage[english]{babel}
\usepackage{amssymb}
\usepackage{epsfig}
\def\lsim{\raisebox{-.4ex}{$\stackrel{<}{\scriptstyle \sim}$\,}}
\def\gsim{\raisebox{-.4ex}{$\stackrel{>}{\scriptstyle \sim}$\,}}

\begin{document}
\newcommand{\A}{{\mathcal{A}}}
\newcommand{\dA}{\delta{\mathcal{A}}}
\newcommand{\Od}{{\cal O}}
\newcommand{\degree}{^\circ}
\newcommand{\K}{\textrm{K}}


\title{Cosmological electromagnetic fields and dark energy}%

\author{Jose Beltr\'an Jim\'enez and Antonio L. Maroto}

\address{Departamento de  F\'{\i}sica Te\'orica I,
Universidad Complutense de Madrid, 28040 Madrid, Spain.}
\date{\today}

\begin{abstract}
We show that the presence of a temporal electromagnetic field on
cosmological scales generates an effective cosmological constant
which can account for the accelerated expansion of the universe.
Primordial electromagnetic quantum fluctuations produced during
electroweak scale inflation could naturally explain the presence
of this field and also the measured value of the dark energy
density. The behavior of the electromagnetic field on cosmological
scales is found to differ from the well studied short-distance
behavior and, in fact, the presence of a non-vanishing
cosmological constant could be signalling the breakdown of gauge
invariance on cosmological scales. The theory is compatible with
all the local gravity tests, and is free from classical or quantum
instabilities. Thus we see that, not only the true nature of dark
energy can be established without resorting to new physics, but
also the value of the cosmological constant finds a natural
explanation in the context of standard inflationary cosmology.
This mechanism could be discriminated from a true cosmological
constant by upcoming observations of CMB anisotropies and large
scale structure.

\end{abstract}

\maketitle

\section{Introduction}

The nature of dark energy, which is believed to be responsible for
the present phase of accelerated expansion of the universe
\cite{SN1,SN2,SN3,SN4},
still remains unknown. Despite its
phenomenological
success, the simplest
description in terms of a cosmological constant ($\Lambda$CDM model)
suffers from an important naturalness problem, since the measured value
of $\Lambda$, corresponding to
$\rho_\Lambda\sim\rho_M\sim (2 \times 10^{-3}$ eV)$^4$,  finds no natural
explanation in the context of known physics. Moreover, the fact that
today matter and
dark energy have comparable contributions to the energy density,
turns out to be difficult
to understand if dark energy is a true cosmological constant. Thus,
the energy density of a
cosmological constant remains  constant throughout the history
of the universe,
whereas those of the rest of components (matter or radiation) grow as we
go back in time.
Then the question arises as to whether it is a {\it coincidence} (or not) that
they have comparable values today when they have differed
by many orders of magnitude in  the past.
Notice also that if $\Lambda$ is a fundamental constant of nature, its scale
(around $10^{-3}$ eV) is more than 30 orders of magnitude smaller than
the natural scale of  gravitation, $G=M_P^{-2}$ with $M_P\sim 10^{19}$ GeV.
On the other hand, if $\Lambda$ is just an effective parametrization of
dark energy, still a proper understanding of the underlying physics would be
needed in order to explain the measured value.

Alternative models have been proposed
in which dark energy is a dynamical component rather than a
cosmological constant.
Such models are usually based on new physics, either in the form of
new  cosmological fields or
 modifications of Einstein's gravity
\cite{quintessence1,quintessence2,quintessence3,quintessence4,
quintessence5}. However, they are generically plagued by
classical or quantum instabilities, fine tuning problems
or inconsistencies with local gravity constraints.

In this paper we explore the possibility of understanding dark energy
from the standard electromagnetic field, without the need of introducing
new physics (previous works on models of dark energy based on
vector fields can be found in \cite{vector1,vector2,vector3,
vector4,vector5,vector6,BeMa,vector7}).
We will show that the behavior of electromagnetic fields on very large
(super-Hubble) scales  differs from the
well studied short-distance (sub-Hubble) behavior. Thus,
on super-Hubble scales, the time component
of the electromagnetic field grows linearly in time in the matter
and radiation eras, giving rise to a
cosmological constant contribution in the electromagnetic energy-momentum
tensor (the potential
gravitational effects of longitudinal
electromagnetic fields were  considered in a different context in \cite{Deser}).
 At late times this contribution becomes dominant giving rise to
the accelerated phase.  As a possible generating mechanism,
we calculate the spectrum of super-Hubble
electromagnetic modes produced during inflation from quantum fluctuations
and find that
the correct value of the dark energy density can be naturally obtained in the
case in which inflation took place at the electroweak scale.

\section{Cosmological electromagnetic fields}

We start by writing the electromagnetic action including a
gauge-fixing term in the presence of gravity:
\begin{eqnarray}
S&=&\int d^4x\sqrt{-g}\left[-\frac{1}{16\pi G}R-\frac{1}{4}F_{\mu
\nu}F^{\mu\nu}+\frac{\lambda}{2}\left(\nabla_\mu
A^\mu\right)^2\right]
\label{action}
\end{eqnarray}
The gauge-fixing term is required in order
to define a consistent quantum theory for the electromagnetic
field \cite{Itzykson}, and we will
see that it plays a fundamental role on large scales.
Still this action preserves a residual gauge symmetry
$A_\mu\rightarrow A_\mu+\partial_\mu \phi$ with $\Box \phi=0$.

Einstein's and electromagnetic equations derived from this
action can be written as:
\begin{eqnarray}
R_{\mu\nu}&-&\frac{1}{2}Rg_{\mu\nu}=8\pi
G\left(T_{\mu\nu}+T^A_{\mu\nu}\right)\\
\nabla_\nu F^{\mu\nu}&+&\lambda\nabla^\mu\nabla_\nu
A^\nu=0\label{fieldeq}
\end{eqnarray}
where $T_{\mu\nu}$ is the energy-momentum tensor for matter and
radiation and $T^A_{\mu\nu}$ is the energy-momentum tensor of the
electromagnetic field. Notice that since
we will be using the covariant Gupta-Bleuler formalism, we
do not a priori impose the Lorentz condition.
The effect of the high conductivity of the
universe in the matter and radiation eras will be discussed below.

We shall first focus on the simplest case of a homogeneous electromagnetic
field (zero mode) in a flat Robertson-Walker background, whose
metric is given by:
\begin{equation}
ds^2=dt^2-a(t)^2\delta_{ij}dx^i dx^j
\end{equation}
In this space-time, equations (\ref{fieldeq}) read:
\begin{eqnarray}
\ddot{A}_0+3H\dot{A}_0+3\dot{H}A_0&=&0\label{eqA0t}\\
\ddot{\vec{A}}+H\dot{\vec{A}}&=&0\label{feqRW}
\end{eqnarray}
with $H=\dot{a}/{a}$  the Hubble parameter.

Notice that (\ref{eqA0t}) implies that the gauge-fixing term  exactly
behaves as a cosmological constant throughout the history of the universe,
irrespective of the background evolution. Indeed, for homogeneous fields
we have:
\begin{eqnarray}
\frac{d}{dt}(\nabla_\mu A^\mu)=\frac{d}{dt}(\dot A_0+3HA_0)=0\label{const}
\end{eqnarray}

We can solve (\ref{eqA0t}) and (\ref{feqRW}) during the radiation and matter
dominated epochs when the Hubble parameter is given by $H=p/t$
with $p=1/2$ for radiation and $p=2/3$ for matter. In such a case
the solutions for (\ref{feqRW}) are:
\begin{eqnarray}
A_0(t)&=&A_0^+t+A_0^-t^{-3p}\label{A0sol}\\
\vec{A}(t)&=&\vec{A}^+t^{1-p}+\vec{A}^-\label{Azsol}
\end{eqnarray}
where $A_{0}^\pm$ and $\vec{A}^\pm$ are constants of integration.
Hence, the growing mode of the temporal component does not depend
on the epoch being always proportional to the cosmic time $t$,
whereas the growing mode of the spatial component evolves as
$t^{1/2}$ during radiation and as $t^{1/3}$ during matter, i.e.
at late times the temporal component will dominate over the
spatial ones.

On the other hand, the $(^0\;_0)$ component of Einstein's
equations adopts the following form:
\begin{eqnarray}
H^2=\frac{8\pi G}{3} \left[\sum_{\alpha=R,M}
\rho_\alpha+\rho_{A_0}+\rho_{\vec{A}}\right] \label{Friedmann}
\end{eqnarray}
where $R, M$ stands for radiation and matter respectively and:
\begin{eqnarray}
\rho_{A_0}&=&\lambda\left(\frac{9}{2}H^2A_0^2+3HA_0\dot{A}_0+\frac{1}{2}
\dot{A}_0^2\label{rhoA0}\right)
\\
\rho_{\vec{A}}&=&\frac{1}{2 a^2}(\dot{\vec{A}})^2
\end{eqnarray}
Notice that we need $\lambda>0$ in order to
have positive energy density for $A_0$. Besides, when inserting
the  solutions (\ref{A0sol}) and (\ref{Azsol})
into these expressions we obtain
that $\rho_{A_0}=\rho_{A_0}^0$, $\rho_{\vec{A}}=\rho_{\vec{A}}^0\,a^{-4}$
and $\nabla_\mu A^\mu=$ const as commented before.
Thus, the field behaves as a cosmological constant
throughout the evolution of the universe since its temporal
component gives rise to a constant energy density  whereas the
energy density corresponding to $\vec{A}$ always decays
as radiation.
 Moreover, this fact
 prevents the generation of a non-negligible anisotropy
which could spoil the highly isotropic  CMB radiation
(see \cite{Barrow} for a more general discussion).
Finally, when the universe is dominated by
the cosmological constant arising from the gauge-fixing term, 
both the Hubble parameter and
$A_0$ become constant
leading therefore to a future de Sitter
universe. Let us emphasize that according to (\ref{const}), $\rho_{A_0}$
always contributes as a cosmological constant. As the observed fraction 
of energy density associated to
a cosmological constant today is $\Omega_\Lambda\simeq 0.7$, we
obtain that the field value today must be $A_0(t_0)\simeq 0.3\,M_P$.

The effects of the high electric conductivity $\sigma$ can
be introduced using the
magneto-hydro\-dynamical approximation and including
on the r.h.s. of
Maxwell's equations the corresponding current term, 
which is given by $J_\mu-J_\nu u^\nu u_\mu=\sigma F_{\mu\nu}u^\nu$
with $u^\mu$ the velocity associated to the
comoving observers. Notice that the 
strict neutrality of the plasma, which is consistent with a  
vanishing  electric field, implies 
$J_\mu u^\mu=0$, and  finally,  
the current can be written as:
$J_\mu=(0,\sigma(\partial_0  A_i-\partial_i A_0))$.  
Notice that  electric
 neutrality also  implies that  conductivity does
not affect the evolution of $A_0(t)$. The infinite conductivity
limit simply eliminates the growing mode of $\vec A(t)$ in
(\ref{Azsol}). The inhomogeneous case, corresponding to $k\neq 0$
modes, will be discussed in next section.

We still need to understand which are the appropriate initial conditions
leading to the  present value of $A_0$. In order to avoid the
cosmic coincidence problem, such initial conditions
 should have been set in a natural way in the
early universe.
In a very interesting work \cite{Arkani},
 it was suggested that the present value
of the dark energy density could be related to physics at the
electroweak scale since $\rho_\Lambda \sim (M_{EW}^2/M_P)^4$, where
$M_{EW}\sim 10^3$ GeV. This relation offers
a hint
on the possible mechanism generating the initial
amplitude of the electromagnetic fluctuations. Indeed, we see that if
 such amplitude is set
by the size of the Hubble horizon at the electroweak era, i.e.
$A_0(t_{EW})^2\sim H_{EW}^2$, then
the correct scale for the dark energy density is obtained.
Thus, using the Friedmann equation, we find $H_{EW}^2\sim M_{EW}^4/M_P^2$, but
according to (\ref{rhoA0}), $\rho_{A_0}\sim H^2A_0^2\sim\;$const.,
so that $\rho_{A_0}\sim H_{EW}^4\sim (M_{EW}^2/M_P)^4$
as commented before.

A possible implementation of this mechanism can take place
during inflation.
Notice that
the typical
scale of the dispersion of quantum field fluctuations on super-Hubble
scales generated
in an inflationary period is precisely set
by the almost constant Hubble parameter during such period $H_I$,
i.e. $\langle A_0^2\rangle\sim H_I^2$ \cite{Linde}.  The
 correct dark energy density can then be
naturally obtained if initial conditions for the
electromagnetic fluctuations are set during an  inflationary
epoch at the scale $M_I\sim M_{EW}$. Let us make
these arguments more precise.

\section{Quantum fluctuations during inflation}

We shall look at the electromagnetic perturbations generated during inflation
 in order to determine its primordial power
spectrum. In this case it is more convenient to use conformal time
$\eta$ defined by means of $dt=ad\eta$ and to introduce the
conformal components of the field $\A_\mu=(aA_0,\vec{A})$. Besides
we shall focus on a single Fourier mode of the vector field with
wave vector $\vec{k}$ and decompose the field in temporal, transverse
and longitudinal components with respect to $\vec{k}$. In this frame,
equations (\ref{fieldeq}) read:
\begin{eqnarray}
\A_{0k}''&-&\left[\frac{k^2}{\lambda}-2{\mathcal{H}}'
+4{\mathcal{H}}^2\right]\A_{0k}
-2ik\left[\frac{1+\lambda}{2\lambda}\A_{\parallel k}'
-{\mathcal{H}}\A_{\parallel k}\right]=0 \label{modes}\\
\vec{\A}_{\perp k}''&+&k^2\vec{\A}_{\perp k }=a\sigma\vec{\A}_{\perp k}'
\nonumber\\
\A_{\parallel k}''&-&k^2\lambda\A_{\parallel
k}-2ik\lambda\left[\frac{1+\lambda}{2\lambda}\A_{0k}'
+{\mathcal{H}}\A_{0k}\right]=a\sigma({\A}_{\parallel k}'-ik\A_{0k})\nonumber
\end{eqnarray}
with $'\equiv\frac{d}{d\eta}$ and ${\mathcal{H}=aH}$ is the Hubble
parameter in conformal time.  He have included for completeness the current
term on the right-hand side as commented before. Notice that once again
the electric neutrality of the
universe implies that the evolution equation for the temporal component
is not modified. During inflation the
electric conductivity of the universe is negligible and this term can
be safely neglected so that in the following we shall set $\sigma=0$.

It is easy to see from equations (\ref{modes})  that the transverse
modes are just plane waves  irrespective of the expansion rate.
On the other hand, the components $\A_{0k}$ and $\A_{\parallel k}$
are coupled to each other even in the absence of gravity. This is
due to the fact that we are working with arbitrary $\lambda$ and
not using the simple Feynman gauge $\lambda=-1$.

Let us first consider quantization in Minkowski space-time,
with ${\cal H}={\cal H}'=0$.
The decomposition in Fourier modes can be
written as follows:
\begin{eqnarray}
\A_{0}&=&\int\frac{d^3\vec{k}}{2k_0(2\pi)^3}
\left[\left(-i\frac{1+\lambda}{1-\lambda}({\bf a}_0+{\bf a}_\parallel)
k\eta
+{\bf a}_0\right)e^{-ikx}\right.\nonumber \\
&+&\left.\left(i\frac{1+\lambda}{1-\lambda}({\bf a}_0^+ +{\bf a}_\parallel^+)k
\eta+{\bf a}_0^+\right)e^{ikx}\right]\nonumber\\
\A_{\parallel}&=&\int\frac{d^3\vec{k}}{2k_0(2\pi)^3}\left[\left(i\frac{1
+\lambda}{1-\lambda}({\bf a}_0+{\bf a}_\parallel)k\eta+{\bf a}_\parallel\right)
e^{-ikx}\right.\nonumber \\
&+&\left.\left(-i\frac{1+\lambda}{1-\lambda}({\bf a}_0^++{\bf a}_\parallel^+)
k\eta+{\bf a}_\parallel^+\right)e^{ikx}\right]\label{solsMink}
\end{eqnarray}
with $k_0=\vert \vec k\vert=k$.
Now, in order to have the canonical commutation rules
$[\A_\mu(t,\vec{x}),\Pi^\nu(t,\vec{y})]=
i\delta^\nu_\mu\delta^{(3)}(\vec{x}-\vec{y})$,
the creation and annihilation
operators appearing in (\ref{solsMink}) should satisfy:
\begin{eqnarray}
\left[{\bf a}_0(\vec{k}),{\bf a}_0^+(\vec{k'})\right]=&&
\frac{1-\lambda}{\lambda}k_0(2\pi)^3\delta^{(3)}(\vec{k}-\vec{k'})\nonumber\\
\left[{\bf a}_\parallel(\vec{k}),{\bf a}_\parallel^+(\vec{k'})\right]=&
-&\frac{1-\lambda}{\lambda}k_0(2\pi)^3\delta^{(3)}(\vec{k}-\vec{k'})
\end{eqnarray}
For simplicity in the following we will take $\lambda=1/3$ so that we
use canonically normalized operators with positive sign for the temporal
component. Notice that this is just the opposite situation
to the usual Feynman gauge. In fact, $\lambda=1/3$ and $\lambda=-1$ are the
only two possible choices with canonical normalizations.
As is well-known \cite{Itzykson}, in order to recover Maxwell's theory,
we need to eliminate the negative norm states
by defining the corresponding restricted Hilbert space. Following the
standard Gupta-Bleuler formalism,
the physical states $|\phi\rangle$  will be those annihilated by the
combination ${\bf a}_0+{\bf a}_\parallel$, that is: $({\bf a}_0(\vec k)
+{\bf a}_\parallel(\vec k))
|\phi\rangle=0$. In Minkowski space-time, only transverse
degrees of freedom contribute  to the expectation
value of the energy density in the
physical states and $\left\langle \phi \vert
T_{00}\vert \phi\right\rangle> 0$ since the contributions
from longitudinal and temporal
modes cancel each other. Thus, as expected, the theory
is free from ghosts.
Notice that in Minkowski space-time, we also get
$\langle \phi\vert\partial_\mu A^\mu\vert\phi\rangle=0$.

Now we can proceed to the quantization in the inflationary epoch.
In order to present the calculational method explicitly,  we assume an
exact
de Sitter phase.  The general quasi-de Sitter results will be given below.
Thus in de-Sitter:  $a=-1/(H\eta)$
and ${\mathcal{H}}=-1/\eta$ with $\eta<0$. The classical solutions of the
corresponding equations are a bit
more complicated, although it is still possible to obtain
analytic expressions:
\begin{eqnarray}
\A_{0k}&=&C_1k\eta e^{-ik\eta}
+\frac{C_2}{k\eta}\left[\frac{1}{2}(1+ik\eta)e^{-ik\eta}-k^2\eta^2e^{ik\eta}E_1(2ik\eta)\right]\nonumber\\
\nonumber\\
\A_{\parallel k}&=&iC_1(1+ik\eta)e^{-ik\eta}
-iC_2\left[\frac{3}{2}e^{-ik\eta}+(1-ik\eta)e^{ik\eta}E_1(2ik\eta)\right]
\end{eqnarray}
where $E_1(x)=\int_1^\infty e^{-tx}/tdt$ is the exponential
integral function. Note that the mode $C_1$ can be gauged
away  by means of a residual gauge transformation.
The sub-Hubble limit ($\vert k\eta\vert \gg 1$) of these solutions reads:
\begin{eqnarray}
\A_{0k}&=&\left(C_1k\eta +iC_2\right)e^{-ik\eta}\nonumber\\
\A_{\parallel
k}&=&\left(-C_1k\eta-iC_2\right)e^{-ik\eta}\label{sub}
\end{eqnarray}
The choice of  adiabatic vacuum \cite{Birrell} is made by matching these
solutions with those obtained in the Minkowski case (\ref{solsMink}) (up to
 sub-leading terms in the sub-Hubble limit). To do this, we
choose the modes $C_1$ and $C_2$ in the following way:
\begin{eqnarray}
C_1&\rightarrow& -\frac{i}{k_0}\left({\bf a}_0
+{\bf a}_\parallel\right)\nonumber\\
C_2&\rightarrow&-\frac{i}{2k_0}{\bf a}_0
\end{eqnarray}

On the other hand, on super-Hubble scales ($\vert k\eta \vert \ll 1$) we have:
\begin{eqnarray}
\A_{0k}&=&\frac{1}{2}C_2(k\eta)^{-1}\nonumber\\
\A_{\parallel
k}&=&iC_1-iC_2\left(\frac{3}{2}-\gamma-\ln(2ik\eta)\right)\label{super}
\end{eqnarray}
with $\gamma$ the Euler's constant. We see that $\A_{0k}\propto a$, which
means that $A_{0k}=a^{-1}\A_{0k}$ is (almost) constant during inflation,
once the mode leaves the horizon.
In the case of quasi-de Sitter slow-roll inflation,
the Hubble parameter reads ${\mathcal{H}}=-1/((1-\varepsilon)\eta)$, where the
slow-roll parameter is defined as $\varepsilon=1/(16\pi G)(V'/V)^2\ll 1$,
with $V$
the inflaton potential. Following the same steps as before,
we obtain the
power spectrum for $A_0$ on super-Hubble scales:
\begin{eqnarray}
{\mathcal{P}}_{A_0}(k)\equiv\frac{k^3}{2\pi^2}
\langle\left|A_{0k}\right|^2\rangle
=
\frac{H_I^2}{16\pi^2}\left[\frac{k}{aH_I}\right]^{n_{A_0}}
\end{eqnarray}
which is almost scale-invariant (as in the scalar field case) since for the
electromagnetic spectral index we obtain $n_{A_0}=-4\varepsilon$.
In a similar way it is possible to obtain the primordial power spectrum
of longitudinal modes on super-Hubble scales:
\begin{eqnarray}
{\mathcal{P}}_{A_\parallel}(k)=\frac{k^2}{16\pi^2 \varepsilon^2}
\left[\frac{k}{aH_I}\right]^{-4\varepsilon}
\end{eqnarray}
If we now compare the power spectra for the conformal fields ${\cal A}_0$
and ${\cal A}_\parallel$ we find that:
\begin{eqnarray}
\frac{{\mathcal{P}}_{{\cal A}_\parallel}(k)}{{\mathcal{P}}_{{\cal A}_0}(k)}=
\frac{1}{\varepsilon^2}\left(\frac{k}{aH_I}\right)^2
\end{eqnarray}
which is negligible on super-Hubble scales, and allows us
to safely ignore  the longitudinal modes on such scales after inflation.

Notice that since $\varepsilon>0$, ${\mathcal{P}}_{A_0}(k)$  is
a red-tilted spectrum which means
that the contribution to $\langle A_0^2\rangle$ from long wavelenghts
dominates over small scales. In particular, provided
inflation lasted for a sufficiently large number of e-folds,
this allows to decompose the fluctuations field at any given time
into a large  homogeneous
contribution (with scales $k< {\cal H}$)
and a small inhomogeneous  perturbation ($k> {\cal H}$),
and therefore we can use
standard perturbation theory around the homogeneous background. Thus,
for the homogeneous part we
get:
\begin{eqnarray}
\langle A_0^2\rangle_{hom}=\int^{k_*}_{k_{min}}\frac{dk}{k}
{\mathcal{P}}_{A_0}(k)
\simeq H_I^2 \frac{e^{-n_{A_0} \tilde N}}{16\pi^2\vert n_{A_0}\vert}
\end{eqnarray}
where $k_*\lsim H_0$, $\tilde N=N_{tot}-N_0$ and
$k_{min}=e^{-\tilde N}H_0$ is set by the
Hubble horizon at the beginning of inflation \cite{Liddle}. Here $N_{tot}$
is the total number
of e-folds of inflation which should not be confused with $N_0$ which
is the number of e-folds  since the time
when the scale $H_0^{-1}$ left the horizon. Typical values for $N_0$ are around
50, whereas generically there is no upper limit to $N_{tot}$.
Thus as expected, up to tilt corrections, $H_I$ sets the scale
for the field dispersion.
\begin{figure}
\vspace{1cm}
\begin{center}{\epsfxsize=10 cm\epsfbox{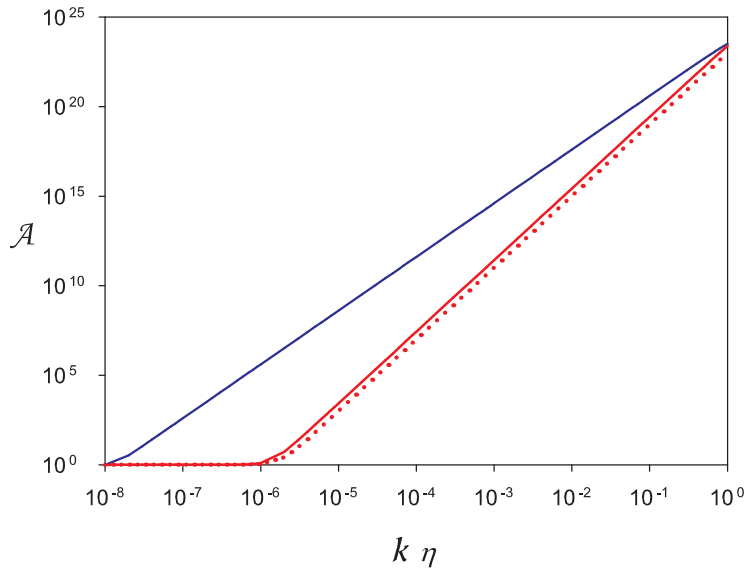}}
\caption{\small Evolution of $\A_{0k}$ and $\A_{\parallel k}$
on super-Hubble scales in the radiation era.
Continuous (dotted) blue lines correspond to
$\A_{0k}$ for infinite (vanishing) conductivity (no difference in the plot).
Continuous (dotted) red lines correspond to
$\A_{\parallel k}$ for infinite (vanishing) conductivity.}
\end{center}
\end{figure}

Once the fluctuations generated during inflation enter the
radiation dominated era, it would be in principle possible that
the high conductivity of the universe could modify their
evolution, spoiling the growing behavior of super-Hubble models
found in (\ref{A0sol}).  However since conductivity does not
affect the temporal equation, the $A_0$ modes are not modified on
super-Hubble scales. In Fig.1 we show the evolution of
super-Hubble temporal and longitudinal modes, both for vanishing
and infinite conductivity. We see that the evolution exactly
corresponds to $\A_{0k}\propto \eta^3$ as expected from
(\ref{A0sol}) in the radiation era, even in the infinite
conductivity case. We see that $\A_{\parallel k}$ is sub-dominant
compared to $\A_{0 k}$, until the modes re-enter the Hubble radius
for $\vert k\eta\vert\simeq 1$. The result is not sensitive to the
change of initial conditions. Thus, the only effect of
the high conductivity is the damping of the electric field,(which is 
consistent with the strict neutrality of the plasma). In
particular, this implies ${\A}_{\parallel k}'=ik\A_{0k}$, which 
corresponds to the field evolution shown in Fig. 1. Let us
emphasize that the vanishing of the electric field does not imply the 
vanishing of the temporal component $\A_{0k}$.

In fact, it is straightforward to show that the value
of $\nabla_\mu A^\mu$ giving rise to the effective cosmological
constant is not affected by the presence of conductivity.
Indeed, Maxwell's equations in the presence of conserved currents read:
\begin{eqnarray}
\nabla_\nu F^{\mu\nu}&+&\lambda\nabla^\mu\nabla_\nu
A^\nu=J^\mu
\label{Maxeq}
\end{eqnarray}
Taking the four-divergence of the equation we get:
 \begin{eqnarray}
\Box(\nabla_\mu A^\mu)=0
\label{Maxeq}
\end{eqnarray}
where we have used current conservation $\nabla_\mu J^\mu=0$. Thus, 
we see that 
the field $\nabla_\mu A^\mu$ evolves as a free scalar field, and
it is therefore constant on super-Hubble scales, independently
of the presence of external currents.

\section{Gauge invariance on cosmological scales}

A remarkable consequence of the  covariant quantization formalism
 in the context of inflationary cosmology (which had not been considered
 previously) is the breaking of gauge
invariance on cosmological scales.
Indeed, in this formalism, the classical energy-momentum
tensor depends on the gauge-fixing term.
However, in Minkowski space-time, when one takes the expectation value of
this object
in a physical state (that belonging to the restricted Hilbert space)
 the gauge
dependence disappears because the contributions from the temporal
and longitudinal
degrees of freedom cancel each other. This is so just because
in Minkowski space-time,
in the restricted Hilbert space, the amplitudes of temporal and
longitudinal mode
solutions are exactly the same (see \cite{Itzykson}). However when
considering the quantization in an expanding universe important
differences arise. At
short distances, i.e. for sub-Hubble modes, it is easy to see from (\ref{sub})
that the same
cancellation takes place, as it should be, and the theory is exactly
the same as in
Minkowski space-time. Therefore the energy density does not depend on
the gauge-fixing
term. Nevertheless, when the modes become super-Hubble, it can be seen
from (\ref{super}), that
the amplitude of the temporal modes grows in time faster than that of
the longitudinal
ones. This spoils the mentioned cancellation and a net energy density
results from the $\lambda$ term.

Notice
that in the covariant formalism the four polarizations are always present.
In Minkows\-ki space-time (or for sub-Hubble modes) only two of them contribute to the
energy density, however on cosmological scales also the temporal one can have
observational consequences. As $A_0$ is also a propagating degree of
freedom,
the gauge-fixing term can be seen in the Gupta-Bleuler formalism as a kinetic term
for it, and therefore the coefficient $\lambda$ can be fixed by the standard
normalization of the creation and annihilation operators.
This effect could not be studied in other (non-covariant) formalisms,
such as Coulomb gauge quantization, since only transverse polarizations would
be present in that case.

To summarize, the presence of the background cosmological electromagnetic
field breaks $U(1)_{EM}$ symmetry on large scales while
preserving local (small-scales) invariance.
This is  analogous to the situation with Lorentz symmetry,
where the presence of matter or radiation in the Universe breaks
global Lorentz invariance, but respecting local transformations.
 In other words, the presence of
a non-vanishing cosmological constant could be signalling
the breakdown of gauge invariance on cosmological scales.
Let us emphasize that this effect is a consequence of the quantization of
 electromagnetic theory in the covariant formalism and,
as discussed above, it does
not modify any  of the physical predictions of Maxwell's theory
for laboratory
experiments or astrophysical observations. As a matter of fact,
the electromagnetic
interaction has not been tested on distance scales larger than
1.3 AU \cite{Nieto}.

\section{Perturbations and consistency}

Despite the fact that the background evolution in the present case
is the same as
in $\Lambda$CDM, the evolution of metric perturbations could
be different, thus offering an observational way of discriminating
between the two models. With this purpose, we have calculated the evolution of metric,
matter density and
electromagnetic perturbations  in the longitudinal gauge
with $\delta g_{00}=2a^2\Phi$, $\delta g_{0i}=a^2S_i$,
$\delta g_{ij}=a^2(2\Psi\delta_{ij}-h_{ij})$,
$\delta=\delta \rho_M/\rho_M$
and taking
$\A_\mu=\A_\mu^{hom}(\eta)+\delta \A_\mu$, where as
commented before the main contribution to $\A_\mu^{hom}(\eta)$
comes from the temporal component. The propagation speeds
of scalar, vector and tensor perturbations are found
to be real and equal to the speed of light, so that the theory is
classically stable. We have also checked that  the theory does
not contain ghosts and it is therefore stable at the quantum level.
On the other hand, using the explicit expressions in \cite{Will} for
the  vector-tensor theory of gravity corresponding to the action in
(\ref{action}),  it is
possible to see that all the parametrized post-Newtonian (PPN) parameters
agree with those of General Relativity,  i.e. the theory is compatible
with all the local gravity constraints for any value
of the homogeneous background electromagnetic field.

\begin{figure}
\vspace{1cm}
\begin{center}{\epsfxsize=9.5 cm\epsfbox{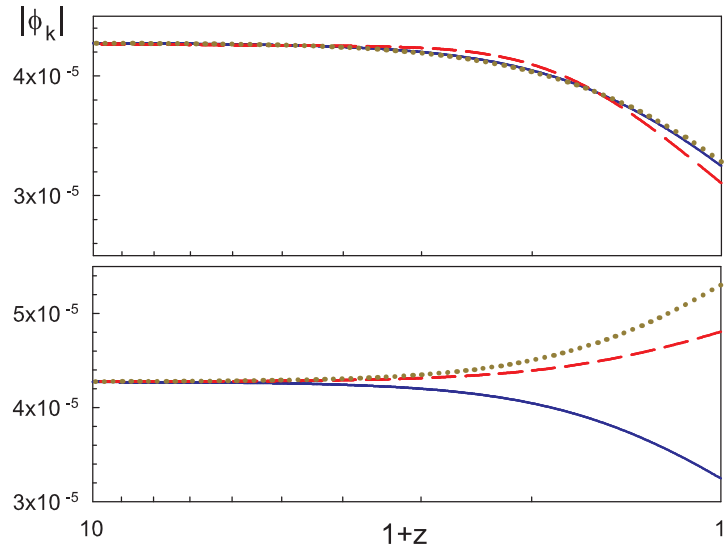}}
\caption{\small Evolution of $\Phi_k$ with $k=3H_0$, corresponding
to the maximum contribution to the ISW effect. Upper(lower) panel
with vanishing(infinite) conductivity.  Continuous blue line for
$\Lambda$CDM, dashed red for $\tilde N\vert n_{A_0}\vert=$12
(upper panel) 18 (lower panel) and
 dotted green for large $\tilde N\vert n_{A_0}\vert$.}
\end{center}
\end{figure}

\begin{figure}
\vspace{1cm}
\begin{center}{\epsfxsize=10 cm\epsfbox{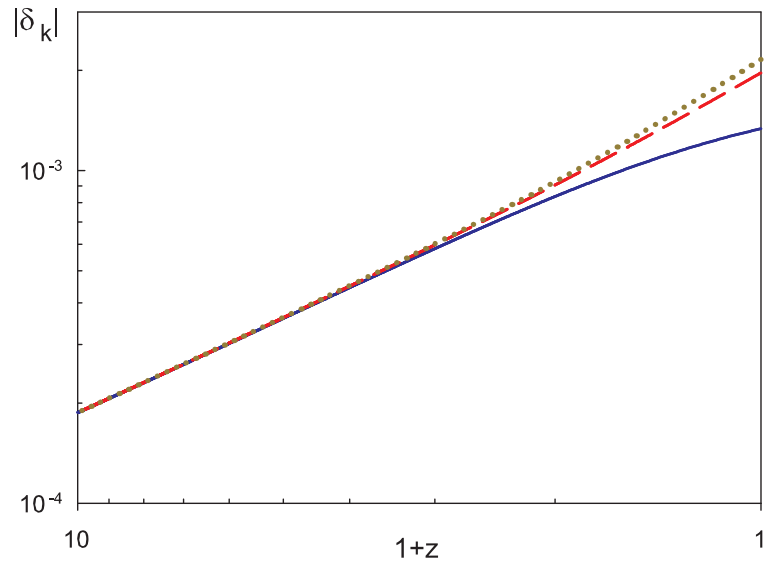}}
\caption{\small Evolution of the matter density contrast
$\delta_k$ with $k=10H_0$ and infinite conductivity (vanishing
conductivity shows no difference with respect to $\Lambda$CDM).
Plotted curves correspond to the same models as in the lower panel
of Fig.2.}
\end{center}
\end{figure}

In Figs 2-3 we plot the
evolution of
scalar perturbations, satisfying $\Phi_k=\Psi_k$
in this theory,
and matter density contrast $\delta_k$, in both, vanishing and
infinite conductivity limits.
We find
that  the only relevant deviations with respect to $\Lambda$CDM
appear on large scales $k\sim H_0$ and that
they depend on the primordial
spectrum of electromagnetic fluctuations. However, for
$\tilde N\vert n_{A_0}\vert \gsim 12$,
the results on the CMB temperature power spectrum and
evolution of density perturbation are compatible with observations.
Taking $M_I\sim T_{RH}\sim 10^2$ GeV in this case,
with $T_{RH}$ the reheating temperature, we obtain the correct value of
the dark
energy density today. In addition, the different evolution of $\Phi_k$
with respect to the $\Lambda$CDM model gives rise to
a possible discriminating  contribution to the
late-time integrated Sachs-Wolfe effect \cite{Turok}.

The presence of large scale electric fields generated by inhomogeneities
in the $A_0$ field opens also the possibility
for the generation of large scale currents which in turn
could contribute to the presence of magnetic fields
with large
coherence scales. This could shed light on the problem of explaining the
origin of cosmological magnetic fields.
Work is in progress in this direction.

\section{Conclusions and discussion}

 We have shown that the present phase of accelerated expansion of
the universe can be explained by the presence of a cosmological
electromagnetic field generated during inflation. This result not
only offers a solution to the problem of establishing the true nature of
dark energy, but also explains the value of the cosmological constant
without resorting to new physics. In this scenario the
fact that matter and dark energy densities {\it coincide} today
is just a consequence of inflation taking place at the electroweak scale.
Such a relatively low inflation scale implies also that no cosmological
gravity wave background is expected to be measurable in future CMB polarization
observations.

Notice also that any vector-tensor theory (not necessarily electromagnetism)
whose low-energy effective
action is given by (\ref{action}) and in which the vector field only interacts
gravitationally with the rest of particles would provide a natural
model for dark energy. In fact all the previous models trying to account
for the cosmic acceleration
are plagued by classical or quantum instabilities, fine tuning problems or
inconsistencies with Solar System experiments. However, in this work we
present, for the first time, an explanation to the cosmic acceleration with
none of the aforementioned problems.

\vspace{0.5cm}

{\em Acknowledgments:} We would like to thank J.D. Barrow,
J.A.R. Cembranos and C. Tamarit for useful comments and suggestions.
This work has been  supported by
DGICYT (Spain) project numbers FPA 2004-02602 and FPA
2005-02327, UCM-Santander PR34/07-15875, CAM/UCM 910309 and
MEC grant BES-2006-12059.


\vspace{0.5cm}
\begin{thebibliography}{99}
\bibitem{SN1} S. Perlmutter et al., {\it Astrophys. J.} {\bf 517}, 565 (1999)
\bibitem{SN2}A.G. Riess et al., {\it Astron. J.} {\bf 116}, 1009 (1998) and
{\bf 117}, 707 (1999)
\bibitem{SN3} D. N. Spergel et al. \emph{Astrophys. J. Suppl.} \textbf{148},
175, (2003) and  astro-ph/0603449
\bibitem{SN4} M. Tegmark et al., {\it Phys. Rev.} {\bf D69}: 103501, (2004).
\bibitem{quintessence1} C. Wetterich, {\it Nucl. Phys.} {\bf B302}, 668 (1988);
\bibitem{quintessence2}R.R. Caldwell, R. Dave and P.J. Steinhardt, {\it Phys. Rev. Lett.}
 {\bf 80}, 1582 (1998)
\bibitem{quintessence3} C. Armendariz-Picon, T. Damour and V. Mukhanov,
{\it Phys. Lett.} {\bf B458}, 209 (1999)
\bibitem{quintessence4} S.M. Carroll, V. Duvvuri, M. Trodden, M.S. Turner,
{\it Phys. Rev.} {\bf D70}: 043528, (2004)
\bibitem{quintessence5} G. Dvali, G. Gabadadze and M. Porrati,
{\it Phys. Lett.} {\bf B485}, 208 (2000)
\bibitem{vector1} V.V. Kiselev, {\it Class. Quant. Grav.}
{\bf 21}: 3323, (2004)
\bibitem{vector2} C. Armendariz-Picon,
{\it JCAP} {\bf 0407}: 007, (2004)
\bibitem{vector3}
C.G. Boehmer and T. Harko, {\it Eur. Phys. J.}
{\bf C50}: 423, (2007)
\bibitem{vector4} M. Novello, et al. {\it Phys. Rev.} {\bf D69}:
127301, (2004)
\bibitem{vector5} T. Koivisto and D.F. Mota, arXiv:0805.4229 [astro-ph]
\bibitem{vector6}H.S. Zhao, {\it Astrophys. J.}, {\bf 671},  L1-L4 (2007).
\bibitem{BeMa} J. Beltr\'an Jim\'enez and A.L. Maroto, {\it Phys. Rev.} {\bf D78}, 063005 (2008)
 and arXiv:0807.2528 [astro-ph]
\bibitem{vector7}K.~Bamba, S.~Nojiri and S.~D.~Odintsov,
  Phys.\ Rev.\  {\bf D77}, 123532 (2008)
\bibitem{Deser} S. Deser, {\it Ann. Inst. Henri Poincar\'e}, {\bf 16}: 79
(1972)
\bibitem{Itzykson} C. Itzykson and J.B. Zuber, {\it Quantum Field Theory},
McGraw-Hill (1980)
\bibitem{Barrow} J.D. Barrow,  {\it Phys. Rev.} {\bf D55}, 7451 (1997);
J.D. Barrow and R. Maartens, {\it Phys. Rev.} {\bf D59}: 043502 (1999)
\bibitem{Arkani}
  N.~Arkani-Hamed, L.~J.~Hall, C.~F.~Kolda and H.~Murayama,
{\it Phys.\ Rev.\ Lett.}\  {\bf 85} (2000) 4434
\bibitem{Linde} A. Linde, {\it Particle physics and inflationary
cosmology}, Harwood Academic Press (1996)
\bibitem{Birrell} N.D. Birrell and P.C.W. Davies, {\it Quantum fields in
curved space}, Cambridge (1982)
\bibitem{Liddle} A.R. Liddle and D.H. Lyth, {\it Cosmological inflation and
large-scale structure}, Cambridge (2000)
\bibitem{Nieto} A.S. Goldhaber and M.M. Nieto, arXiv:0809.1003 [hep-ph]
\bibitem{Will} C. Will, {\it Theory and experiment in gravitational physics},
Cambridge University Press, (1993)
\bibitem{Turok} R.G. Crittenden and N. Turok, {\it Phys.\ Rev.\ Lett.}\  {\bf 76}
(1996) 575

\end{thebibliography}
\end{document}